# Electron-Electron Interaction and Weak Antilocalization Effect in a Transition Metal Dichalcogenide Superconductor

*Chushan Li, Mebrouka Boubeche, Lingyong Zeng, Yi Ji, Qixuan Li, Donghui Guo, Qizhong Zhu, Dingyong Zhong, Huixia Luo, Huichao Wang\**

C. Li, Y. Ji, Q. Li, D. Guo, D. Zhong, H. Wang
School of Physics, Sun Yat-sen University, Guangzhou 510275, China
Email: wanghch26@mail.sysu.edu.cn

M. Boubeche, L. Zeng, H. Luo
School of Materials Science and Engineering, Key Lab of Polymer Composite & Functional Materials, Guangzhou Key Laboratory of Flexible Electronic Materials and Wearable Devices, Sun Yat-sen University, Guangzhou 510275, China

Q. Zhu
Guangdong Provincial Key Laboratory of Quantum Engineering and Quantum Materials, School of Physics and Telecommunication Engineering, South China Normal University, Guangzhou 510006, China
Guangdong-Hong Kong Joint Laboratory of Quantum Matter, Frontier Research Institute for Physics, South China Normal University, Guangzhou 510006, China

D. Zhong, H. Luo
State Key Laboratory of Optoelectronic Materials and Technologies, Sun Yat-sen University 510275 Guangzhou, China

In disordered transition-metal dichalcogenide (TMD) superconductor, both the strong spin-orbit coupling (SOC) and disorder show remarkable effects on superconductivity. However, the features of SOC and disorder were rarely detected directly. Here we report the quantum transport behaviors arising from the interplay of SOC and disorder in the TMD superconductor 1T-NbSeTe. Before entering the superconducting state, the single crystal at low temperature shows a resistivity upturn, which is $T^{1/2}$ dependent and insensitive to the applied magnetic fields. The magnetoresistance (MR) at low temperatures shows a $H^{1/2}$ dependence at high magnetic fields. The characteristics are in good agreement with the electron-electron interaction (EEI) in a disordered conductor. In addition, the upturn changes and MR at low magnetic fields suggest the contribution of weak antilocalization (WAL) effect arising from the strong SOC in the material. Moreover, the quantitative analyses of the transport features in different samples imply anomalous disorder-enhanced superconductivity that needs to be further understood. The results reveal the disorder enhanced EEI and the strong SOC induced WAL effect in 1T-NbSeTe, which illustrate the resistivity minimum in the widely studied doped superconductors. The work also provides insights into the disorder effect on the superconductivity.

1. **Introduction**

Transition-metal dichalcogenides have become recently excellent platforms for investigations on various electronic orders due to their fruitful structural and electronic properties.[1-6] Within the materials, charge/spin density-wave, Mott insulator, excitonic insulator, spin liquid state and superconducting order can be found. Moreover, the characteristic layered structure of TMDs makes them be easily tuned by pressure, doping, intercalation and layer thickness. Thus, TMDs have triggered tremendous interest in many fields including the materials science and physics. In superconducting TMDs, the interplay between superconductivity and other electronic orders is an important topic.[7] Generally, the delicate balance between different orders can be tipped by artificially introduced disorder, and then the underlying interactions may be revealed. Among different methods, elemental substitution is a useful probe widely used to introduce disorder, and a lot of focus was put on the detect of the interplay.[8,9] In addition, the interplay of disorder and interactions is also a fruitful area of investigation.[10,11] The effects of disorder on superconductivity are elusive, involving the interplay of superconductivity, localization and the evolution of Coulomb interactions. In fact, the disorder induced EEI and quantum interference effects are also manifested in the normal phase while they have been rarely investigated. In order to fully understand the role of disorder in these materials, it is necessary to draw attention to the quantum correction to the normal state.

The strong intrinsic SOC is one of the most intriguing properties of TMDs.[12-16] In disordered TMD electronic systems, the interplay of strong SOC and disorder gives rise to different transport behaviors. According to Altshulter-Aronov model,[11] the EEI is expected to be enhanced and gives a quantum correction to the classical Drude conductivity at low temperatures. The EEI contribution is isotropic and does not show any dependence on the magnetic field. In addition, the EEI in disordered structures is usually accompanied by a weak localization (WL) effect. The WL arises from the constructive quantum interferences of time-reversal scattering trajectories remaining the phase coherence. With a phase shift induced by strong SOC, the destructive quantum interference induces a so-called WAL effect.[17] The WL/WAL increases/decreases the localization of the electrons, and thus induces a positive/negative quantum correction term to the resistivity. Since the quantum interference effects are sensitive to the applied magnetic fields, both WL and WAL can be suppressed by external magnetic fields and show corresponding negative and positive MR, respectively. These quantum correction effects on the conductance are generally observed at low temperatures serving as an efficient probe to detect quantum transport in various attractive materials.[18-23]

Recently, the strong SOC effects on TMD superconductors have attracted a lot of attention.[13,24-28] The TMD 2H-NbSe$_2$ is a conventional superconductor with bulk $T_c$ of 7.3 K, the correlation between the superconductivity, charge density wave, Ising superconductivity, strong SOC and its multiband character is under active discussion.[29-31] On the other hand, the distorted 1T TMD superconductor NbTe$_2$ shows signature of SOC induced $p$-wave paring.[32] However, there are few reports about the related quantum interference phenomena in these superconductors. The partial substitution of Se with Te atoms in 2H-NbSe$_2$ enables a new TMD superconductor with pure 1T-NbSeTe phase.[33] The random occupation of Se/Te sites in the system introduces non-negligible disorder makes it a promising platform to probe the SOC and disorder in the TMDs through the electron transport study. Here we experimentally revealed the WAL and EEI in the 1T-NbSeTe. At low temperatures, the resistivity upturn showing $T^{1/2}$ dependence and the MR showing $H^{1/2}$ dependence indicate the dominant EEI contribution. In addition, the resistivity response at low magnetic fields suggests the coexisting WAL effect, providing insight into the resistivity minimum in doped conductors at low temperatures. Moreover, the observed anomalous phenomenon of disorder enhanced superconductivity is discussed. The work is helpful to understand the interplay of SOC, disorder and superconductivity.

2. Method

The single crystals of 1T-NbSeTe were grown via the chemical vapor transport method using I$_2$ as a transport agent as reported.[33] The obtained crystals show a layered structure with metallic silver grey in color. The composition and the structure of the crystals were investigated by energy dispersive x-ray spectroscopy, scanning electron microscopy, x-ray diffraction (XRD) and high-resolution transmission electron microscopy, which reveal that the obtained samples are 1T-NbSeTe crystallized in the trigonal structure with space group P-3m1.[33] The composition of three elements is close to 1:1:1 and the Se/Te occupies sites randomly. Inset of Figure 1a shows the XRD pattern of a typical sample (sample 1, s1), in which the (00n) peaks indicate that the $c$ axis is along the normal direction of the blade-shaped crystal plane. The electronic transport measurements were conducted in a 14T-PPMS using the standard four-electrode method.

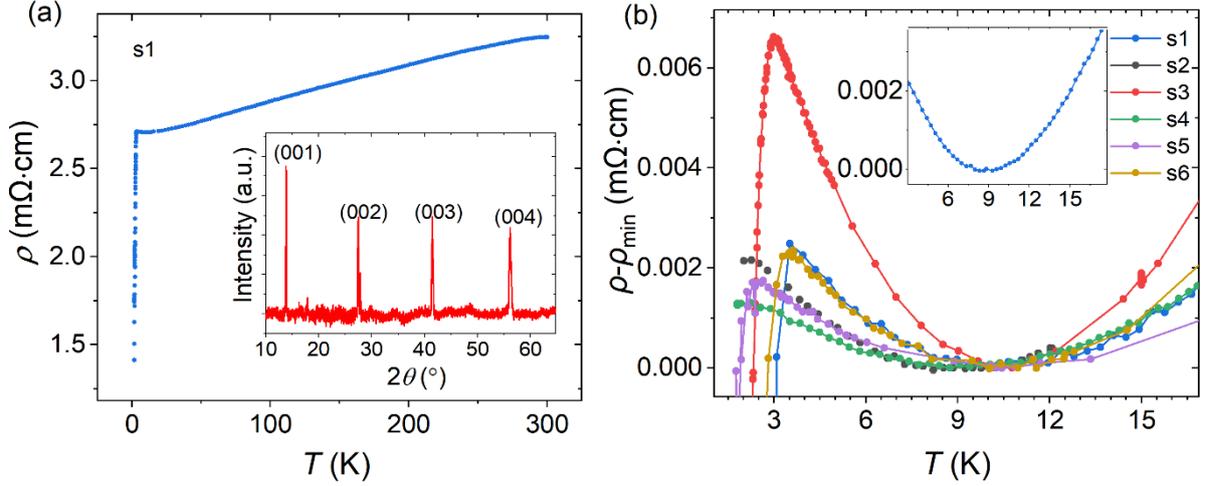

**Figure 1.** Temperature dependent resistivity of the 1T-NbSeTe single crystals. (a) The resistivity deceases with decreasing temperature showing metallic feature at high temperature followed by a sharp reduce at low temperature around 3 K implying the superconducting transition. Inset: XRD pattern for single crystal revealing the top plane is (001). (b) A resistivity upturn appears before the superconducting transition inducing a resistivity minimum (inset).

### 3. Results and discussion

The temperature dependent resistivity ($\rho$) of the single crystal 1T-NbSeTe is shown in Figure 1(a). As the temperature is decreased from 300 K, the resistivity decreases exhibiting the typical metallic behavior. At low temperature of about 3 K, a sharp decrease of the resistivity indicates the superconducting transition, consistent with previous report on the material.[33] The zero resistance was not observed here because the lower temperature to millikelvin is not available in the measurement facility. Nevertheless, the bulk superconductivity in the 1T-NbSeTe have been demonstrated by the previous results including the zero resistance and the heat capacity characterization.[33] In addition, we observe a resistivity minimum in the single crystals at low temperature around 10 K. Inset of Figure 1b shows the enlarged behavior around the resistivity minimum of s1. Similar results are also observed in other 1T-NbSeTe samples. To scale the resistivity change in different samples, the vertical axis values in Figure 1b and inset are displayed as $\rho-\rho_{min}$, i.e., the resistivity subtracted the corresponding resistivity minimum ($\rho_{min}$).

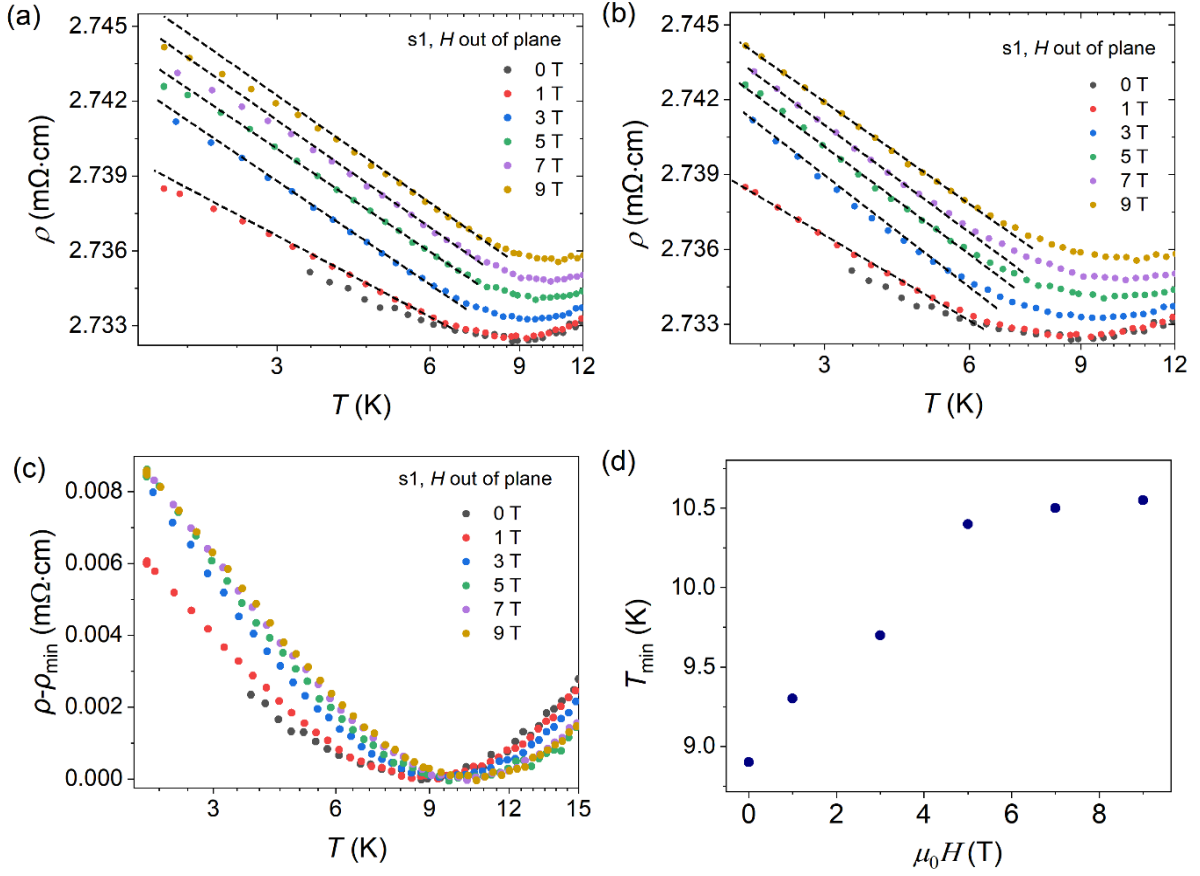

**Figure 2.** Resistivity minimum in 1T-NbSeTe at applied perpendicular magnetic fields. (a) The resistivity vs. temperature in a logarithmic scale. The low temperature data deviate from the dashed line. (b) The resistivity vs. temperature in the scale of $T^{1/2}$. The dashed lines are guides to the eye. (c) Resistivity subtracted the minimum value vs. temperature. (d) Characteristic temperatures $T_{min}$ of the resistivity minimum vs. applied magnetic fields. The $T_{min}$ increases at low fields and remains almost unchanged at high fields.

The resistivity upturn appears at low temperatures and the correction to the resistivity $(\rho-\rho_{min})/\rho_{min}$ is estimated to be in the order of 0.3% at 2 K, which suggest the quantum correction as the origin. Related mechanisms are the Kondo effect,[34] the WL effect[17] and the EEI.[11] Owing to the enhanced thermal fluctuations and decoherence mechanism at high temperatures, the quantum corrections to conductivity are generally observed at low temperatures. The Kondo effect arises from the exchange interaction between itinerant conduction electrons and localized spin impurities. The absence of any magnetic impurities excludes the Kondo effect as the possible underlying mechanism. In the 1T-NbSeTe single crystals, the random Se/Te sites introduce non-negligible disorder. It is expected that the EEI can be enhanced by disorder, and the EEI contribution was observed in the resistivity of many disordered metallic systems at low temperature.[18-22] Thus, we need to examine the role of the EEI in the resistivity upturn.

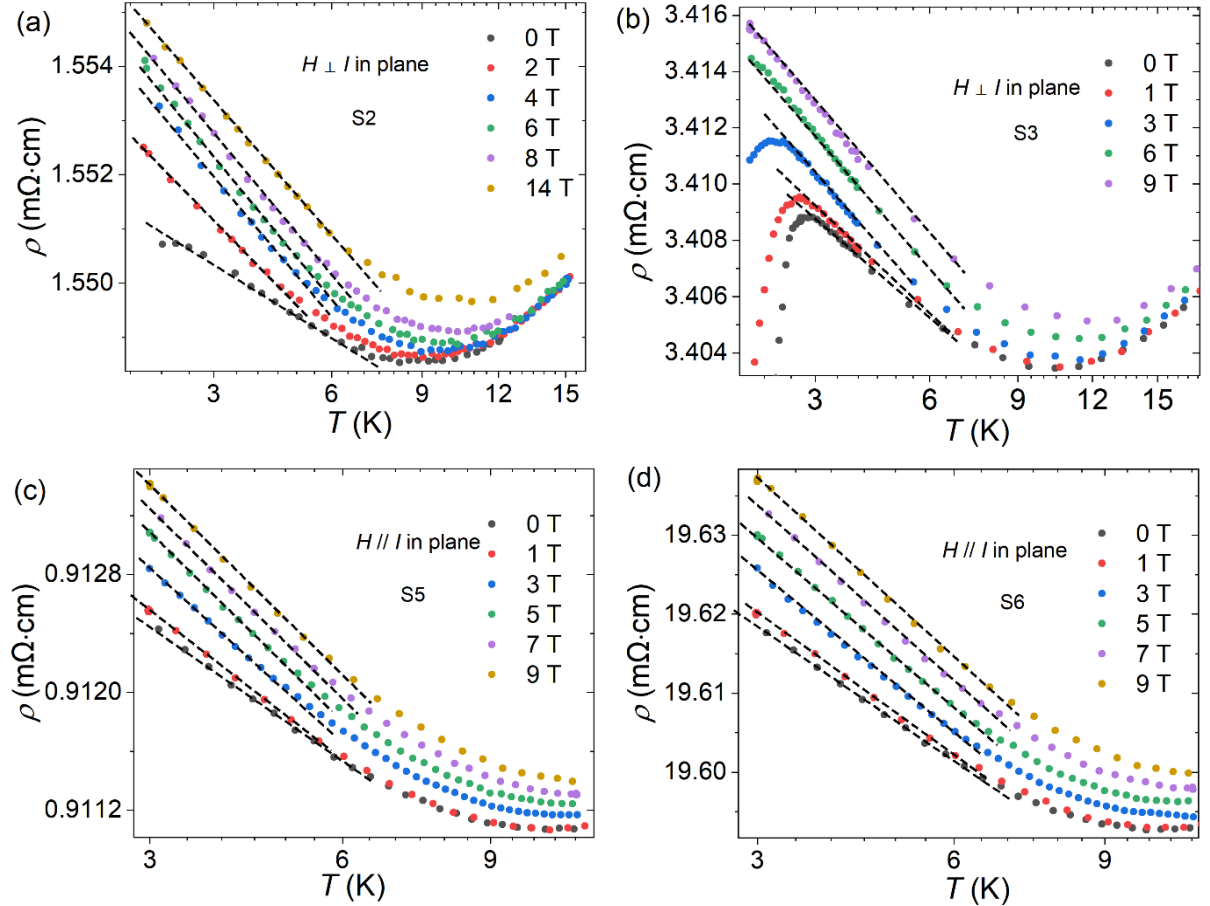

**Figure 3.** Resistivity upturn in 1T-NbSeTe at parallel magnetic fields. Resistivity vs. temperature of (a) s2 and (b) s3 in the form of $T^{1/2}$ when the in-plane magnetic field is perpendicular to the current. Resistivity at low temperatures ($T^{1/2}$) in (c) s5 and (d) s6 when the in-plane field is parallel to the current. The isotropic $T^{1/2}$ dependence reveals the dominant role of EEI in the resistivity upturn. The dashed lines are guides to the eye.

According to the theoretical prediction, EEI in the presence of disorder produces a quantum correction term to resistivity with $\Delta\rho_{EEI} \propto -lnT$ in the two-dimensional (2D) systems and $\Delta\rho_{EEI} \propto -T^{1/2}$ in the three-dimensional (3D) systems.[11] Here, the bulk 1T-NbSeTe is expected to exhibit the 3D nature. On the other hand, the TMD 1T-NbSeTe hosting the layered structure featuring the quasi-2D dimensionality. To find the most suitable formula, we plot the resistivity upturn versus temperature both in log-scale and in the form of $T^{1/2}$, which are shown in Figure 2a and 2b, respectively. It can be observed that the upturn shows better match with the $T^{1/2}$ dependence while it always deviates from the ln$T$ dependence at very low temperatures. Furthermore, we plot the relative change of resistivity ($\rho-\rho_{min}$) in Figure 2c. It is found that the high magnetic field data above 3 T almost coincide. Meanwhile, the temperature of the resistivity minimum ($T_{min}$) also shows little change above 3 T (Figure 2d). Thus, the resistivity upturn at high magnetic fields is insensitive to the magnetic fields, which is in agreement with the 3D EEI picture.

The isotropic feature is also studied to further confirm the contribution from EEI. As shown in Figure 3, the resistivity upturn always keeps the $T^{1/2}$ dependence independent of the direction of the magnetic field. The nearly constant slope at high magnetic fields is consistent with the feature in Figure 2c, revealing the dominant EEI at low temperatures and high magnetic fields. In the meanwhile, the EEI manifests itself in the MR. In the Altshuler-Aronov model, the EEI induced MR has a $H^{1/2}$ dependence and is isotropic to H insensitive to the orientation.[11] We plot the MR curves of different NbSeTe samples in Figure 4. Obviously, the MR at low temperatures shows clear $H^{1/2}$ dependence that is independent of the magnetic field direction. Besides, the MR sharply decreases when the temperature is above the $T_{min}$, suggesting the correlation of the enhanced MR and the resistive upturn regime. Thus, the experimental results are well consistent with that in 3D disordered conductors the EEI interaction dominates resulting in the quantum correction to the classical resistivity.

It is worth noting that the MR deviates from the $H^{1/2}$ dependence at low magnetic fields. In addition, the resistively upturn at lower fields shows obvious changes. The results suggest other coexisting mechanism at low temperatures. As shown in inset of Figure 4a, the MR dip around zero field suggests the appearance of WAL,[20,35] which is consistent with the resistivity upturn change at lower fields. As introduced, the WAL induces negative correction term to the resistivity and can be suppressed by magnetic fields. At zero magnetic field, the WAL and EEI coexist in fact though the dominated EEI induces the obvious $T^{1/2}$ dependence. At lower fields, the WAL is suppressed and the resistivity upturn is enhanced. The larger magnetic field induces the higher upturn until the WAL contribution is almost eliminated and the EEI contribution stays unchanged at high magnetic fields.

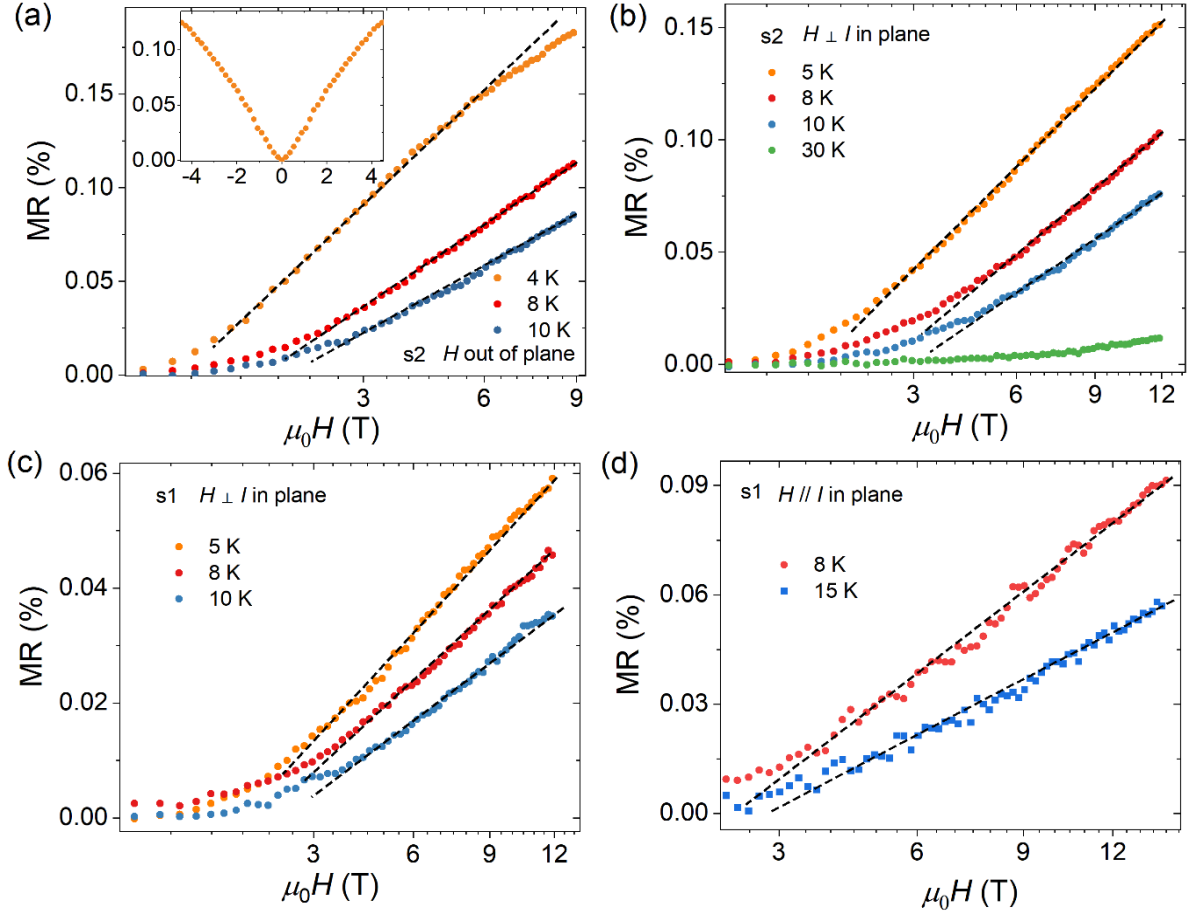

**Figure 4.** The MR at low temperatures vs. magnetic field in the form of $H^{1/2}$. (a) The out of plane MR. (b) (c) The in-plane transverse MR. (d) The in-plane longitudinal MR. The MR at the temperatures which belong to the resistivity upturn regime always show the $H^{1/2}$ dependence. The dashed lines are guides to the eye.

In a system, the quantum interference behavior is crucially characterized by two times scales:[17,36] the dephasing time $\tau_\phi$ and the spin-flip time $\tau_{SO}$. In the regime $\tau_\phi \gg \tau_{SO}$, the spin-orbit scattering is significant.[17] The frequent spin flips make that the quantum interference between the time reversed trajectories is destructive, leading to the occurrence of WAL and a positive MR. In the regime where $\tau_\phi$ is comparable to $\tau_{SO}$ or $\tau_\phi < \tau_{SO}$, the spin-orbit scattering effect is weak and the scattering process barely affects the spin orientation. In such case, negative MR occurs with increasing magnetic field as a WL feature. The observation of WAL rather than WL effect demonstrates the strong SOC in the 3D 1T-NbSeTe system.

We characterized the quantum interference effects and the EEI quantitatively using the phenomenological equation:

$$\Delta\rho_{xx} = \rho_a + aT^{1/2} + bT^2 + cT^{1/3} \qquad \text{(Eq. 1)}$$

where the first and second terms characterize the 3D EEI contribution. The third term is related to the scattering mechanism at relatively higher temperatures above the $T_{min}$. Generally, the electron-phonon scattering at the higher temperatures would contribute to $\sim T^3$ or $\sim T^5$ term to the resistivity of a metallic system. However, as shown in the Figure 5a, the resistivity of 1T-NbSeTe at 15-50 K shows $T^2$ dependence, which indicates that the EE scattering (or Baber scattering) is more important than the electron-phonon scattering at this temperature regime.[23,37] This mechanism is consistent with Fermi liquid theory, which predicts $T^2$ dependence of Baber scattering in metals at low temperature. It is noted that the EE scattering in the regime dominates the origin of the resistivity, which is different from the EEI contributing to a very small quantum correction to the classical residual resistivity below $T_{min}$. The 3D WAL produces $\Delta\rho \propto T^{-p/2}$, where $p$ is defined by the $T$ dependence of the phase coherence time with $p=3$ for phonon scattering and $p=2/3$ for EE scattering.[20] Since the EE scattering dominates here, the WAL contribution is termed as $cT^{1/3}$.

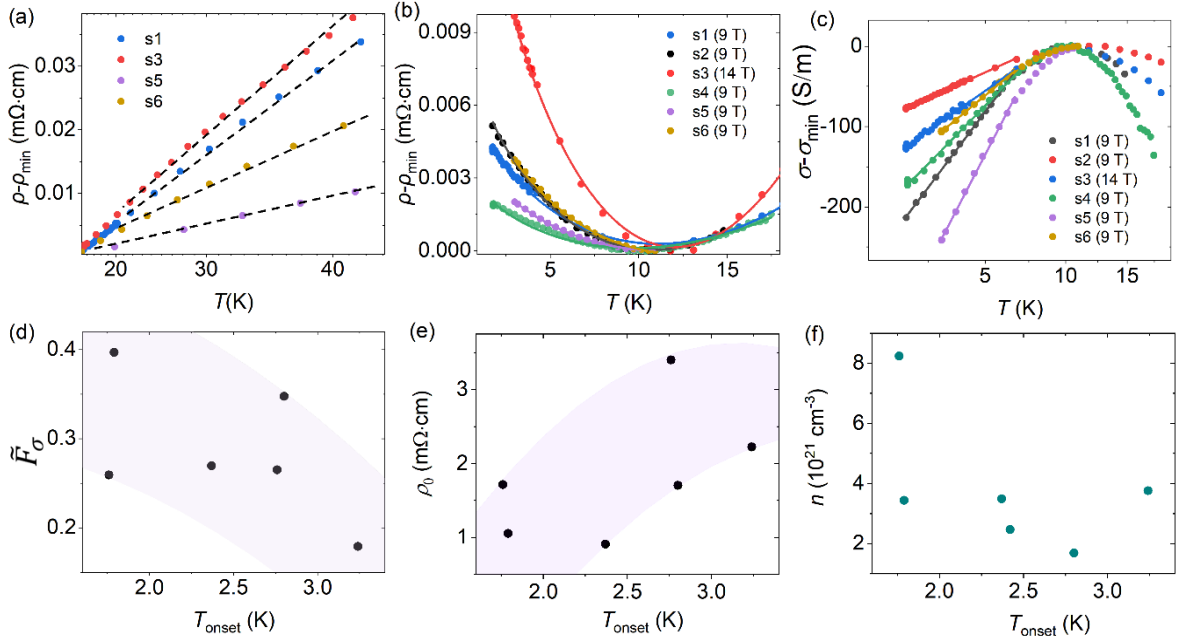

**Figure 5.** Disorder effect on superconductivity. (a) Resistivity vs. temperature in the form of $T^2$ up to 50 K suggests the dominated EE scattering. (b) The resistivity upturn at high magnetic fields fitted by $\Delta\rho_{xx} = \rho_a + aT^{1/2} + bT^2$. (c) The conductivities vs. temperature are fitted by $\delta\sigma_{EEI}(T) = \frac{e^2}{\hbar}\frac{1}{4\pi^2}\frac{1.3}{\sqrt{2D}}(\frac{4}{3} - \frac{3}{2}\tilde{F}_\sigma)\sqrt{T}$. (d) The screening factors $\tilde{F}_\sigma$ shows negative correlation with onset $T_c$. (e) The residual resistivity vs. onset $T_c$. (f) The carrier density vs. onset $T_c$. The results reveal anomalous disorder-enhanced superconductivity.

Since there are four parameters in Equation 1, the result of direct fit without constraints may be not in line with the actual condition of the sample. Thus, the coefficient $a$ and $b$ are estimated by the fit of resistivity upturn at high magnetic fields using the simplified equation of $\Delta\rho_{xx} = \rho_a + aT^{1/2} + bT^2$. The parameter $b$ obtained is comparable to the linear fit results at relatively higher temperatures above $T_{min}$. Then, the zero-field resistivity at low temperatures is fitted by the Equation 1 with fixed $a$ and $b$ values as shown in Figure 5b. The positive value of $c$ further confirms the contribution of WAL effect. It is worth noting here that if the value of $a$ or $b$ is not fixed, the fit may produce a negative $c$, which would induce opposite conclusion for the quantum interference effect in the system.

According to the theory,[11] $\delta\sigma_{EEI}(T) = \frac{e^2}{\hbar}\frac{1}{4\pi^2}\frac{1.3}{\sqrt{2D}}(\frac{4}{3} - \frac{3}{2}\tilde{F}_\sigma)\sqrt{T}$ where $\tilde{F}_\sigma$ is a number characteristic of the Coulomb interaction. In addition, the field dependence of EEI induced MR is $\delta\sigma_{EEI}(H) = -\frac{e^2}{\hbar}\frac{\tilde{F}_\sigma}{4\pi^2}\frac{\sqrt{T}}{\sqrt{2D}}(\sqrt{h} - 1.3)$ where $h = g\mu_B H/kT \gg 1$. Using the theoretical fit of $T^{1/2}$ dependent conductivity (Figure 5c) and the $H^{1/2}$ dependent magnetoconductivity at low temperatures, the screening factor $\tilde{F}_\sigma$ and the diffusive coefficient $D$ are estimated assuming $g = 2$. Within Thomas-Fermi theory, the $\tilde{F}_\sigma$ is between 0 and 1, and the larger $\tilde{F}_\sigma$ means stronger screening. The obtained $\tilde{F}_\sigma$ between 0 and 1 is reasonable indicating disorder weakened screening effect. The diffusive coefficient $D$ is about 10 cm$^2$s$^{-1}$, which is comparable to the report in NbSe$_2$.[38]

To understand the differences between samples with different onset superconducting temperature $T_c$(onset), the $\tilde{F}_\sigma$ vs. $T_c$ for each sample are shown in Figure 5d, respectively. Since the smaller $\tilde{F}_\sigma$ indicates more disorder, the enhanced $T_c$ suggests a disorder enhanced superconductivity feature. The anomalous phenomenon was also manifested by the resistivity vs. $T_c$ in Figure 5e, where a larger resistivity denoting more disorder corresponds to a larger $T_c$. Nonmagnetic disorder was initially believed to show no effect on superconductivity ("Anderson theorem"), while the later Anderson localization theory reveals that disorder suppresses superconductivity and even leads to superconductor-to-insulator transition.[39-43] The disorder was considered to renormalize the EEI and enhance the Coulomb repulsion, which breaks the Cooper pair and thus destroys a superconducting state. However, the anomalous enhancement of superconductivity by disorder has been recently observed in low-dimensional superconductors. For example, the disorder enhanced superconductivity in 2H-NbSe$_2$ was attributed to the suppression of the competing charge density wave.[30] When the Coulomb interaction is screened or unusually weak, the disorder also enhances the superconducting transition temperature originating from the multifractality of electron wave functions leading to an increase of the effective attraction. The enhanced superconductivity in monolayer $NbSe_2$ was attributed to the multifractal superconducting state.[44, 45] Similar results were also observed in $Na_{2-\delta}Mo_6Se_6$, in which an intrinsic screening of long-range Coulomb repulsion in disordered quasi-1D superconductor may play an important role.[46] In the 3D NbSeTe crystals, no other competing electronic order state is observed, and the Coulomb interaction is weakly screened as shown in the text. Thus, it is surprising to find the disorder enhanced $T_c$ in the 3D crystals. The disorder-engineered enhancement of $T_c$ in monolayer $TaS_2$ was ascribed to the increment of carrier density[47] while we did not find similar correlation of the carrier density and $T_c$ (Figure 5f). The underlying origin is elusive now. More experimental and theoretical efforts are needed to clarify the mechanism.

4. Conclusion

In summary, the EEI and the WAL effect are demonstrated in the TMD superconductor 1T-NbSeTe. In the system, the resistivity upturn at low temperatures shows characteristics of EEI including the insensitivity to magnetic fields, isotropic nature and the obvious $T^{1/2}$ dependence. The MR dependence on $H^{1/2}$ also matches well with the EEI mechanism. In addition, the resistivity response to lower magnetic fields reveals the WAL effect due to the strong SOC in the material. Moreover, the quantitative analyses of different samples suggest an anomalous enhancement of superconductivity by disorder, which calls for further understanding. The strong electron correlations and SOC can produce rich phases of matter, e.g., spin liquid, ferromagnetism, Mott insulating, in TMD with 1T phase especially in the monolayer.[29,48-51] It would be interesting to study the thickness dependence and the monolayer 1T-NbSeTe single crystal in the future.

**Acknowledgements**
We acknowledge J.W., Y.X. and M.W. for helpful discussions. The work is supported by the National Natural Science Foundation of China (Grant No. 12004441, 92165204, 11922415, 11774434, 12004118), the Hundreds of Talents program of Sun Yat-sen University and the Fundamental Research Funds for the Central Universities (No. 202lqntd27). Q.Z. acknowledges the support from the Guangdong Basic and Applied Basic Research Foundation (Grants No. 2020A1515110228 and No. 2021A1515010212), and the Science and Technology Program of Guangzhou (Grant No. 2019050001).

**Conflict of Interest**

The authors declare no conflict of interest.

**Data Availability**

The data that support the findings of this study are available from the corresponding authors upon reasonable request.

**Keywords**

quantum transport, disorder-enhanced superconductivity, spin-orbit coupling, electron-electron interaction, resistivity minimum